\documentstyle[aps,preprint,psfig]{revtex}
\begin{document}
\renewcommand{\thesection}{\arabic{section}}
\renewcommand{\thetable}{\arabic{table}}
\tighten
\title{A Wannier-function-based {\em ab initio} 
Hartree-Fock study of polyethylene}
\author{Alok Shukla\cite{email}, Michael Dolg} 
\address{Max-Planck-Institut f\"ur
Physik komplexer Systeme,     N\"othnitzer Stra{\ss}e 38 
D-01187 Dresden, Germany}
\author{Hermann Stoll} \address{Institut f\"ur Theoretische Chemie,
Universit\"at Stuttgart, D-70550 Stuttgart, Germany}

\maketitle

\begin{abstract}
In the present letter, we report the extension of our Wannier-function-based
{\em ab initio} Hartree-Fock approach---meant originally for three-dimensional 
crystalline insulators---to deal with quasi-one-dimensional periodic
systems such as polymers. The system studied is  all-{\em transoid} polyethylene,
and results on optimized lattice parameters, cohesive
energy and the band structure utilizing 6-31G** basis sets are presented.
Our results are also shown to be in excellent
agreement with those obtained with traditional Bloch-orbital-based
approaches.
\end{abstract}
\pacs{ }
\section{INTRODUCTION}
\label{intro}

Polyethylene, with its rather simple geometric and electronic structure, 
can be characterized as a prototype of all insulating polymers. Perhaps
for that reason, and certainly because of its technological importance, 
it is one of the most extensively studied systems,
both experimentally, and theoretically. Its geometrical structure
has been investigated in a number of experiments both by
x-ray diffraction techniques~\cite{pe-exp1,pe-exp2,pe-exp3}, and by
the neutron-scattering method~\cite{pe-exp4}. The electronic structure
of polyethylene has been investigated in photoconduction and photoemission
experiments resulting in the determination of its band gap~\cite{pe-band},
and in angle-resolved photoemission experiments leading to the measurement
of its valence-band structure~\cite{pe-vband1}.
Theoretically it has been investigated by a number of semiempirical
methods~\cite{pe-semp1,pe-semp2,pe-semp3,
pe-semp7,pe-semp8}, by the {\em ab initio} crystal-orbital Hartree-Fock (HF) 
method~\cite{pe-karp,pe-teramae,pe-teramae2,pe-karp2,pe-andre}, and by
subsequent many-body perturbation theory (MBPT) based correlation 
methods~\cite{pe-suhai1,pe-suhai2,pe-sun}. Correlation effects have also 
been accounted for in studies using density-functional
theory~\cite{pe-falk,pe-hsu,pe-lev,pe-ladik}.

Recently we have proposed an approach to study the electronic structure
of periodic insulators in terms of Wannier functions~\cite{shukla1},
with the long-term goal of a wavefunction-based treatment of electron
correlation effects in three-dimensional (3D) crystalline solids. The
approach has been made much more efficient as compared to its original 
implementation, and has since been applied, at the HF level, to study the 
electronic structure, and various related properties, of
a variety of 3D ionic insulators~\cite{shukla2,albrecht,shukla3}.
As a further improvement of the approach, we have extended it to deal
with quasi-one-dimensional (1D) periodic systems such as polymers, as
demonstrated in the present letter 
by means of an all-electron study of all-{\em transoid} polyethylene.
The details of the approach pertaining to its implementation, along with 
its applications to an infinite LiH chain and
all-{\em trans} polyacetylene, are reported elsewhere~\cite{shukla4}. 

As is evident from the results of 
refs.~\cite{pe-suhai1,pe-suhai2,pe-sun}, impressive successes
can be achieved in the {\em ab initio} treatment of electron-correlation 
effects for the case of 1D systems such as polymers,
even within Bloch-orbital-based approaches.  However, generalization
to a 3D solid is far from trivial because, for such a system, owing to
the higher dimensionality of its Brillouin zone, the sum over the virtual
states in the k-space will involve a huge number of terms. Moreover, the 
larger degree of degeneracy
generally associated with Bloch orbitals of a 3D crystal (band
crossing), will make such an approach even more difficult to implement. 
This is the reason behind our belief that Wannier functions offer the most
natural language, in which to formulate the {\em ab initio} 
treatment of electron-correlation effects, as their utility  is not 
restricted to 
systems of any particular dimensionality. And which electrons will contribute
to important correlation effects can be decided almost intuitively, as
electrons localized on Wannier functions which are far-apart, will
interact only weakly. However, being a relatively new approach, it is
important for us to check its applicability on a variety of sytems
even at the HF level. Since a large body of high-quality 
theoretical and experimental data already exists on polyethylene, any
new theoretical study of the compound can be judged against it.
This is the reason that we present a HF study of all-{\em transoid} 
polyethylene in the present letter, utilizing the, relatively large, 
6-31G** basis set employing polarization functions. In order to be 
absolutely certain about the
correctness of our method, we also used the Bloch-orbital-based 
CRYSTAL program~\cite{crystalprog} to study the same compound employing
the identical 6-31G** basis set, and those results are also presented for
comparison. We note that Wannier functions have been employed in earlier
{\em ab initio} studies of electron correlation effects in 
polymers~\cite{ladik}. However, unlike the past studies where the
Wannier functions were obtained by localizing the Bloch orbitals, the present
work leads to direct determination of Wannier functions of the polymers,
without invoking the Bloch orbitals.

The remainder of the paper is organized as follows. In Section \ref{theory}, 
we briefly describe the theory with particular emphasis on the treatment of the
Coulomb lattice sums which differs from the Ewald-summation based approach
adopted for our earlier studies on 3D crystals. 
In section \ref{results}, we present the results
of our calculations which include optimized geometry, cohesive energy, band
structure, and contour plots of some of the Wannier functions. 
Finally, section \ref{conclusion} contains our conclusions.
\section{THEORY}
\label{theory}
In our previous papers~\cite{shukla1,shukla2} we showed that, for a 
crystalline insulator with $2n_c$ electrons per unit cell, one can obtain
its $n_c$ restricted-Hartree-Fock(RHF) Wannier functions localized in
the reference unit cell,  
$\{ |\alpha\rangle; \alpha =1,n_{c} \}$,
by solving the equation
\begin{equation}
( T + U
 +   \sum_{\beta} (2 J_{\beta}-  K_{\beta})   
+\sum_{k \in{\cal N}} \sum_{\gamma} \lambda_{\gamma}^{k} 
|\gamma({\bf R}_{k})\rangle
\langle\gamma({\bf R}_{k})| ) |\alpha\rangle
 = \epsilon_{\alpha} |\alpha\rangle
\mbox{,}
\label{eq-hff1}         
\end{equation}
where $T$ represents the kinetic-energy operator, $U$ represents
the interaction of the electrons of reference cell  with the nuclei
of the whole of the crystal while $J_{\beta}$, $K_{\beta}$  are the
Coulomb and the exchange operators defined as 
\begin{equation}
\left.
 \begin{array}{lll}
 J_{\beta}|\alpha\rangle & = & \sum_{j} \langle\beta({\bf R}_{j})|\frac{1}{r_{12}}|
\beta({\bf R}_{j})\rangle|\alpha\rangle \\  
 K_{\beta}|\alpha\rangle & = & \sum_{j} \langle\beta({\bf R}_{j})|\frac{1}{r_{12}}|\alpha\rangle
|
\beta({\bf R}_{j})\rangle \\  
\end{array}
 \right\}  \mbox{,} \label{eq-jk} 
\end{equation}
The first three terms of Eq.(\ref{eq-hff1}) constitute the canonical Hartree-Fock
operator, while the last term is a projection
operator which makes the orbitals localized in the reference cell orthogonal to those 
localized in the unit cells in the immediate neighborhood of the reference
cell,
in the limit of infinitely high shift parameters $\lambda_{\gamma}^{k}$'s. These
neighborhood unit cells, whose origins are labeled by lattice vectors
${\bf R}_{k}$, collectively define an ``orthogonality region'' and 
are denoted by ${\cal N}$. The 
projection operators along with the shift
parameters play the role of a localizing potential in the Fock matrix, and 
once self-consistency has been achieved, the occupied eigenvectors of 
Eq.(\ref{eq-hff1})  are localized in the reference cell, and are 
orthogonal to the 
orbitals of ${\cal N}$---thus making them Wannier 
functions~\cite{shukla1,shukla2}.
The size of ${\cal N}$ can be denoted by specifying the number
$N$ of nearest neighbors that are included in
${\cal N}$. For example, $N=3$ shall imply that ${\cal N}$ contains up to
third-nearest neighbors of the reference cell, and so on.
From the Wannier functions localized in the reference cell 
(cf. Eq. (\ref{eq-hff1})) one can obtain the corresponding orbitals localized
in any other unit cell by the expression
\begin{equation}
|\alpha({\bf R}_{i})\rangle = {\cal T} ({\bf R}_{i}) 
|\alpha(0)\rangle \mbox{,}
\label{eq-trsym}
\end{equation}
where $|\alpha(0)\rangle$ represents a Wannier orbital localized in
the reference unit cell assumed to be located at the origin while  $|\alpha({\bf R}_{i})\rangle$ is the corresponding orbital
of the $i$-th unit cell located at lattice vector ${\bf R}_{i}$, 
and the corresponding translation
is induced by the operator ${\cal T} ({\bf R}_{i})$.
Thus Eqs. (\ref{eq-hff1}) and (\ref{eq-trsym}) provide us with the complete HF
solution of the infinite crystal.

We have computer implemented the approach outlined above within a 
linear-combination of atomic orbital (LCAO) approach, utilizing Gaussian
lobe-type basis functions~\cite{wannier}.
The nontrivial aspect of the program mainly involves the evaluation of
the infinite lattice sums contained on the right hand side of 
Eq. (\ref{eq-jk}). The evaluation of exchange lattice sums can be performed
entirely in the real-space for insulators quite inexpensively,  
utilizing the rapidly decaying nature of both the lattice sums, as well as
that of the single-particle density matrix. However, the Coulomb lattice
sums are divergent on their own. The contribution of the Coulomb interaction
to the Fock matrix converges only when
the electron-repulsion lattice sums are appropriately combined with the
lattice sums involving electron-nucleus attraction. For the 3D systems
studied studied earlier, we used an Ewald-summation based approach
to perform the lattice sums involved in the 
Coulomb series~\cite{shukla1,shukla2,albrecht,shukla3}. However,  for the
treatment of quasi 1D systems such as polymers, we have adopted an entirely
real-space based approach towards the evaluation of the Coulomb series,
the details of which have been presented elsewhere~\cite{shukla4}.
 For the sake of completeness
we will briefly outline the salient features of the approach here as well.
In this approach, all the matrix elements needed to represent the 
canonical HF operator included 
in Eq. (\ref{eq-hff1}) in the LCAO form can be generated from the 
translationally invariant skeleton of the corresponding 
Fock matrix~\cite{shukla4}
\begin{eqnarray}
 F_{pq}({\bf t}_{pq}) & = & \langle p({\bf t}_{pq})|T| q(0)\rangle 
-\sum_{j=-M}^{M} \sum_{A}^{\mbox{atoms}}
\langle p({\bf t}_{pq}) | \frac{Z_{A}}{|{\bf r} - {\bf R}_{j} -{\bf r}_A|} |
q(0)\rangle \nonumber\\ 
                      &   & + 
2\sum_{j=-M}^{M} \sum_{r,s} \sum_{{\bf t}_{rs}} \langle 
p({\bf t}_{pq})) \:r({\bf t}_{rs}+{\bf R}_{j}) |\frac{1}{r_{12}}
|q(0)  \: s({\bf R}_{j})\rangle D_{rs}({\bf t}_{rs}) \nonumber \\
                      &   & - \sum_{k} \sum_{r,s} \sum_{{\bf t}_{rs}} \langle 
p({\bf t}_{pq})) \: s({\bf R}_{k}) |\frac{1}{r_{12}}
|r({\bf t}_{rs}+{\bf R}_{k}) \: q(0)   \rangle D_{rs}({\bf t}_{rs})
\; \mbox{.} 
\label{fpq}
\end{eqnarray}
The notation used in the equation above is consistent with the
one introduced in our earlier work~\cite{shukla2} where, e.g., 
$| p({\bf t}_{pq})\rangle$ denotes the $p$-th
basis function located in the unit cell labeled by lattice vector 
${\bf t}_{pq}$ and $| q(0)\rangle$ denotes the $q$-th basis function
located in the reference cell. Additionally,
${\bf R}_{j}$, ${\bf R}_{k}$, ${\bf t}_{pq}$, and ${\bf t}_{rs}$
represent (arbitrary) translation vectors of the lattice, and 
$D_{rs}({\bf t}_{rs})$ represents the elements of the one-particle
density matrix of the infinite system assuming that $r$-th function is
located in the cell ${\bf t}_{rs}$ while the $s$-th function is in the
reference cell.
The second and the third elements of the equation 
represent the electron-nucleus and electron-electron interaction
parts of the Coulomb series, respectively. The  lattice
sums representing both types of contributions to the Coulomb series are 
denoted by the sums over index $j$.
 The Coulomb contribution to the Fock matrix element
displayed above is brought to convergence by choosing to terminate both the 
lattice sums at the same and a ``sufficiently-large'' value of the index $j$.
We choose to denote this value by variable $M$, which clearly implies
the inclusion of the Coulomb interaction of the reference cell electrons with
those in up to its $M$-th nearest neighbors.
As a consequence of the fact that the unit cell is electrically neutral,
the divergences inherent in the two Coulomb contributions are equal
and opposite in nature and should cancel themselves for large values
of $M$. In order to obtain the correct value of the total energy 
per unit cell, the nucleus-nucleus repulsion energy is
also computed by restricting the corresponding sum to $M$ nearest neighbors
of the reference cell. The last term of Eq.\ (\ref{fpq}) represents 
the contribution of the exchange interaction.
The corresponding lattice sums represented by
index $k$ converge rapidly in the real space with increasing value of
$k$. The lattice sums 
corresponding to the cell index ${\bf t}_{rs}$ also converge fast
for insulating systems because of the rapidly decaying nature of the
corresponding density matrix elements $D_{rs}({\bf t}_{rs})$, with 
increasing value of $|{\bf t}_{rs}|$~\cite{crystalprog}.
The present, entirely real-space-based approach to the infinite lattice
sums is quite common in polymer studies, and can be found in the
literature dating back to the seventies. An excellent review of the
general real-spaced-based approaches can be found, e.g., in the work
of Delhalle et al.~\cite{delhalle}.

In the theory described above there are two free parameters viz., $N$ which
represents the size of the orthogonality region of the Wannier functions
of the reference cell, and $M$ which represents the range of the Coulomb
interaction included in the Fock matrix. In another paper~\cite{shukla4}
 we have presented a detailed study of the influence of these parameters on the
total energy per unit cell for the infinite LiH chain and 
all-{\em trans} polyacetylene. In all the cases we found that $N=3$ and
$M=75$ yielded total energies converged up to $\approx$ 0.1 mHartree.
In the present study we have used the values $N=5$ and $M=85$.

The band structure presented in this work was obtained using the approach 
outlined in our previous
work~\cite{albrecht}. This is achieved by performing a Fourier transform
on the final converged real-space Fock matrix  (cf. Eq. (\ref{fpq}))
to obtain its reciprocal-space ($k$-space) representation. Then the 
corresponding Fock matrix is diagonalized at 
different $k$ points to obtain the band structure. 

\section{CALCULATIONS AND RESULTS}
\label{results}
\begin{figure}
\centerline{\psfig{figure=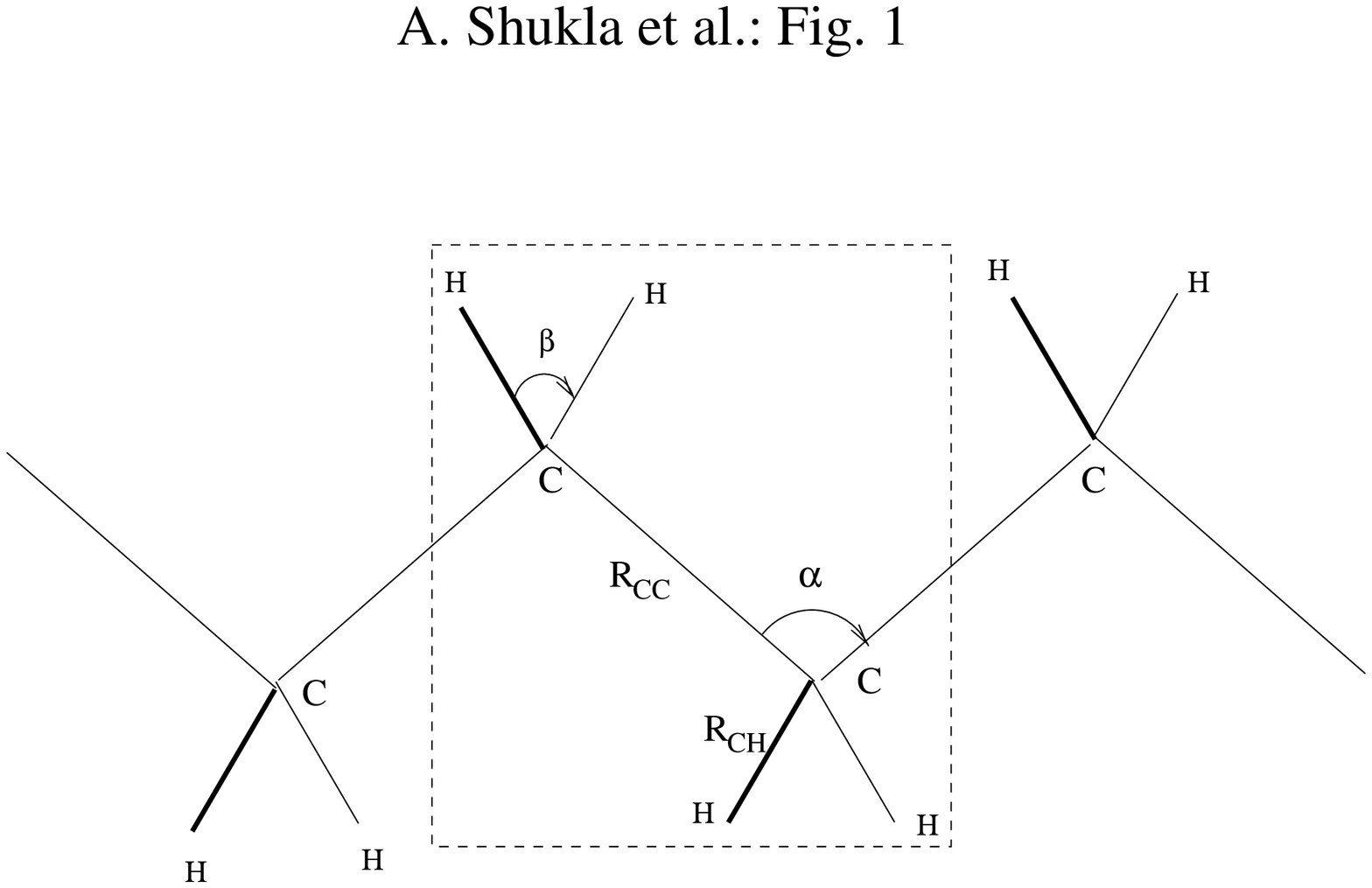,width=13cm,angle=-90}}
\caption{Structure of all-{\em transoid} polyethylene as considered in the present
work. The polymer is assumed to be oriented along the $x$ direction. CC bonds
are in the $xy$ plane while the CH bonds are in the $yz$ plane. Bonds included 
in the reference cell in the calculations
are enclosed in the dashed box; the two CC bonds crossing the borderline are
translationally equivalent, and only one of them is included in the reference cell.}
\label{fig-pe} 
\end{figure}

In this section we present the results of calculations performed on
polyethylene in its all-{\em transoid} geometry.
 The polymer was modeled as a single chain oriented along the x-axis,
as shown in Fig. \ref{fig-pe}.
Since our program is not yet able to utilize the point-group symmetry,
the reference unit cell was assumed to consist of a single C$_2$H$_4$
unit, as against a CH$_2$ unit if the use of the point-group symmetry were
possible. The bonds included in the reference cell during the calculations
are also indicated in Fig. \ref{fig-pe}.
There are four geometrical parameters involved in the structure of
all-{\em transoid} polyethylene namely, carbon-carbon bond length ($R_{CC}$),
carbon-carbon-carbon bond angle ($\alpha$), carbon-hydrogen bond
length ($R_{CH}$) and hydrogen-carbon-hydrogen bond angle($\beta$).
In the present calculations all four parameters were optimized.

We adopted a 6-31G** basis set, the same basis set which we also
used to study all-{\em trans} polyacetylene in 
our previous work~\cite{shukla4}. The polarization functions used
in the basis set consisted of one p-type exponent of 0.75 Bohr$^{-2}$ on 
hydrogen and a single d-type exponent of 0.55 Bohr$^{-2}$ on carbon.
Thus the basis set on carbon was [3s,2p,1d] type while the one on hydrogen
was of [2s,1p] type.
In our calculations, of course, we used the lobe representation of the
corresponding p- and d-type Cartesian basis functions, while the true
Cartesian basis functions were used in the calculations performed with
the CRYSTAL program. For the six atoms included in the reference cell
the present choice of the basis set amounted to 48 basis
functions per unit cell. Since we chose to orthogonalize the Wannier
functions of the reference cell to those residing in its up to
fifth-nearest neighbor cells, we had to include the basis functions
of those cells so as to be able to satisfy the orthogonality condition. 
This increases the total number of basis functions associated with
the reference cell Wannier functions to 528, which, of course, amounts
to an enormous increase in the dimension of the Fock 
matrix (cf. Eq. (\ref{eq-hff1})) to be diagonalized. However, 
the integral evaluation time, which dominates the calculations,
is unaffected by this
proliferation in the basis set size associated with the Wannier functions.
The reason for that is that the one- and two-electron integrals are
evaluated with full regard to the translational symmetry of the
problem~\cite{shukla2,shukla4}, and therefore depend only on the number
of basis functions per unit cell.

Our results on the optimized geometry and the total energy per unit cell
along with the corresponding cohesive energy 
obtained with the 6-31G** basis set are 
presented in table \ref{tab-pef} which also  compares them to the
calculations performed by us---employing the same basis set---with
the CRYSTAL program. As far as the comparison with the works of other
authors  is 
concerned, we will restrict it only to calculations in which 
the {\em ab initio} crystal HF approach was adopted.
To the best of our knowledge, only Karpfen~\cite{pe-karp} and 
Teramae et al.~\cite{pe-teramae} have performed full geometry optimization  
for all-{\em transoid} polyethylene within an {\em ab initio} crystal 
Hartree-Fock approach.
 These authors, of course, employed a Bloch-orbital-based formalism and their 
results are also summarized in
table \ref{tab-pef}. Karpfen~\cite{pe-karp} used both the STO-3G basis 
set as well as an extended basis set employing a [4s,2p] set for carbon and 
a [2s] set for hydrogen. Teramae et al. used only the STO-3G set in their
calculations. Thus, polarization functions were used in neither
of the papers. It is clear from the table that the agreement between the
results obtained with our approach, and the ones with the CRYSTAL program,
is very close. The agreement between our results and those of other
authors is generally satisfactory, i.e.,
the deviations are within the limits of uncertainties
associated with quantum-chemical calculations, except for the carbon-carbon
bond length ($R_{CC}$) which is about 0.03 $\AA$
shorter than the best values reported by Karpfen~\cite{pe-karp} and 
Teramae et al.~\cite{pe-teramae}.
A possible reason is the use of polarization functions in our calculations.
Experimental results on polyethylene are available from x-ray scattering
experiments~\cite{pe-exp1,pe-exp2,pe-exp3} as well as from neutron-scattering
experiments~\cite{pe-exp4}, and are summarized in table \ref{tab-pef}. 
As far as the comparison with the experiments is concerned our results
for all parameters, except for the carbon-hydrogen bond length ($R_{CH}$),
are in good agreement with the experimental results obtained with
the x-ray scattering techniques~\cite{pe-exp1,pe-exp2,pe-exp3}. $R_{CH}$
in the present calculations is overestimated by about 0.03 $\AA$ compared
to the x-ray scattering results. However, for this parameter, results of
other authors also differ by at least that amount. Thus, in our opinion,
this discrepency may be due to electron-correlation effects.
We also present results for the cohesive energy per CH$_2$ unit obtained
from ours as well as CRYSTAL calculations. 
For that purpose
we used Hartree-Fock reference energies for carbon and hydrogen of
-37.677838 a.u. and -0.498233 a.u., respectively. These energies were obtained
by performing atomic HF calculations employing the same 6-31G basis set as
used in the polymer calculations. Our results for the cohesive energy are
in excellent agreement with the one obtained with the CRYSTAL program, 
however, we could not locate any other
theoretical or experimental results on the cohesive energy of polyethylene.

\begin{figure}
\centerline{\psfig{figure=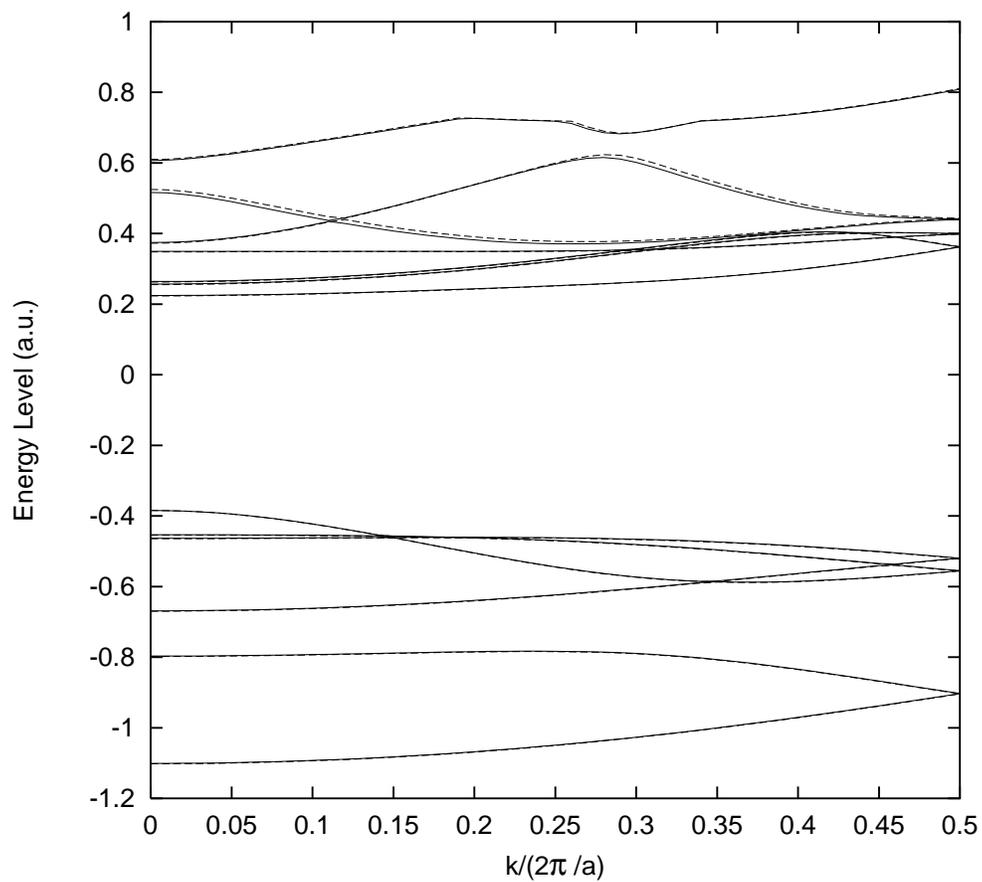,width=13cm,angle=-90}}
\caption{Band structure of polyethylene obtained using our approach 
(solid lines) 
compared to that obtained using the CRYSTAL program (dashed lines). The experimental
geometry (x-ray scattering)~\protect\cite{pe-exp1,pe-exp2,pe-exp3} and a 6-31G** basis set were used in both cases. 
The close agreement between the two sets of bands is obvious.}
\label{fig-band} 
\end{figure}

In Fig. \ref{fig-band} we present the band structure of 
all-{\em transoid} polyethylene computed with the 6-31G** basis set determined
by our approach as well as the one computed using the CRYSTAL program.
Although, in what follows, 
we have not compared our HF band structure with those presented in the
works of other authors~\cite{pe-karp,pe-teramae2}, even a 
cursory comparison reveals that the qualitative features of all the 
Hartree-Fock energy bands are essentially the same.
Our band structure was computed with $N=5$ and $M=100$ while in the CRYSTAL
calculations one hundred $k$ points were used to perform the integration
over the Brillouin zone~\cite{crystalprog}.
 Since our
results for the geometry were in better agreement with the experimental one 
determined  with the x-ray scattering techniques, we chose to evaluate the
band structure for the corresponding geometrical  
parameters~\cite{pe-exp1,pe-exp2,pe-exp3}. Six highest occupied bands along
with the seven lowest unoccupied bands are presented in Fig. \ref{fig-band}.
The absolute values of the band energies naturally 
differed somewhat owing to the different treatment of the Coulomb series in the
two approaches. Therefore we shifted all the CRYSTAL band energies so that 
the tops of the valence bands obtained from the two sets of
calculations coincided.
Clearly the agreement between our band structure and the one obtained
with the CRYSTAL program is very good. The value of the direct band gap 
(at $k= 0$ point) is 
0.6088 a.u. (16.57 eV)  obtained with our approach is in excellent agreement with
the corresponding CRYSTAL value of 0.6077 a.u. (16.54 eV). We believe that
the small differences between our band structure and the one computed
using the CRYSTAL program
are mainly due to the use of lobe functions in our approach to approximate
the Cartesian basis functions used by the CRYSTAL program. 
The value of the direct 
gap for all-{\em transoid} polyethylene was measured to be 8.8 eV
by Less et al.~\cite{pe-band}. Therefore, as is
generally the case with HF bands, the band gap of polyethylene 
is overestimated by a factor of two, pointing to the importance of
the electron correlation effects.  With our approach
the highest point of the valence band occurs at $\approx 10.46$ eV which,
according to the Koopmans theorem, can be directly compared to the ionization
potential of the compound. The experimental value of the ionization potential
of polyethylene is reported to be in the range 9.6--9.8 eV~\cite{pe-vband1}.
Although the agreement between the HF value and the experimental
one is better for the ionization potential as compared to the band gap,
the differences still point to the importance of electron-correlation
effects.  Sun et al.~\cite{pe-sun} recently studied the influence of
electron correlation effects on the valence band of polyethylene using
a Bloch-orbital-based second-order many-body perturbation theory. 
The improvement over the HF results which they observed 
testifies to the general belief that electron correlation effects are
essential to describe the quasi-particle properties in such systems.

\begin{figure}
\centerline{\psfig{figure=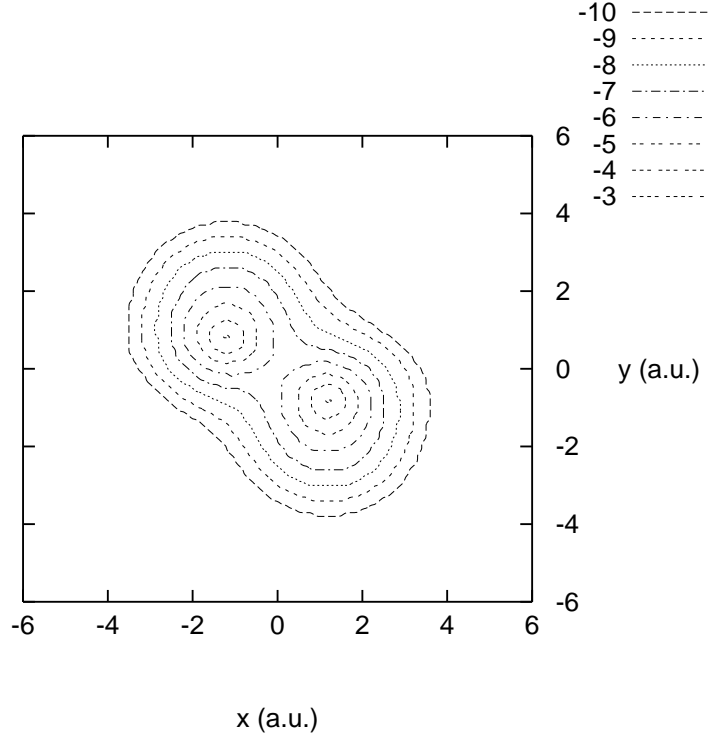,width=13cm,angle=-90}}
\caption{Contour plots of the charge density of the ${\pi}$-type 
valence Wannier function of the reference cell. Contours are plotted
in the $xy$ plane with $z=0.25$ a.u. ($x$ is the axis of the polymer). The
magnitude of the contours is on
a natural logarithmic scale. The two carbon atoms of the 
unit cell are located at the positions $(\mp 1.2, \pm 0.81,0.0)$ a.u.,
while the four hydrogen atoms are at locations  $(\mp 1.2, \pm 2.01, \pm 1.62)$
a.u., approximately. The orbital clearly has a bonding character with respect
to both the carbon-carbon as well as the carbon-hydrogen bond.
The rapidly decaying strength of the contours testifies to the localized nature
of the Wannier function.}
\label{fig-pi1} 
\end{figure}

\begin{figure}
\centerline{\psfig{figure=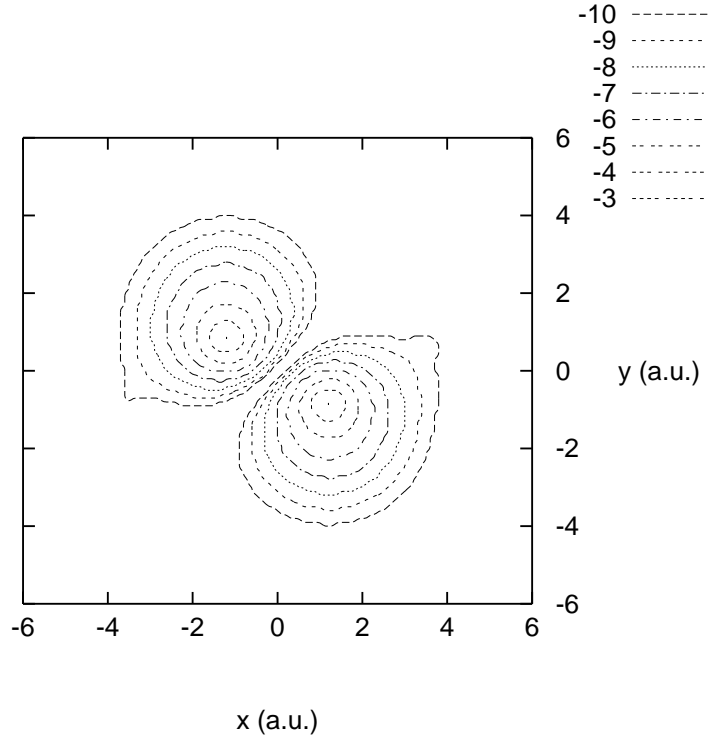,width=13cm,angle=-90}}
\caption{Contour plots of the charge density of the second ${\pi}$-type 
valence Wannier function of the reference cell. The geometrical information
is same as in the caption of Fig. \protect\ref{fig-pi1}.
This contour plot clearly indicates an antibonding
character of the Wannier function with respect to the two carbon atoms of the
unit cell, and a bonding character with respect to the carbon-hydrogen bonds.
Some weaker contours can be seen to reach the carbon
atoms belonging to the nearest-neighbor unit cells.}
\label{fig-pi2} 
\end{figure}

One can obtain an intuitive understanding of the chemical bonding in the 
system by examining the Wannier functions. Such views are presented in
Figs. \ref{fig-pi1} and \ref{fig-pi2} which display  the charge densities
of the Wannier functions corresponding to the 
bonds of the unit cell, which are antisymmetric ($\pi$-like)  
under the reflection about the $xy$ plane. 
The contour
plots correspond to the charge density associated with the corresponding Wannier
function in the $xy$ plane with $z=0.25$ atomic units.
These orbitals were evaluated at the experimental geometry used
also to compute the band structure. Fig. \ref{fig-pi1} clearly implies
a Wannier function which has  a bonding character with respect to the
two carbon atoms of the unit cells, while the second $\pi$-like  
shown in Fig. \ref{fig-pi2}
displays an antibonding character. Both the Wannier functions possess
bonding characters with respect to the C-H bonds of the unit cell.
From the contour plots the localized nature of the orbitals is quite obvious.
The $\sigma$-type Wannier functions (not presented here), although still
localized within the unit cell, are mixtures of both the traditional C-C
and the C-H bonds.

\section{CONCLUSIONS AND FUTURE DIRECTIONS}
\label{conclusion}
 In conclusion, an {\em ab initio} Wannier-function-based Hartree-Fock 
approach developed originally to treat infinite 3D crystalline systems 
has been applied to a 1D system, namely, polyethylene. The main difference 
as compared to the case of 3D systems has been an entirely real-space based 
treatment of the Coulomb series. The approach yields results
which are in excellent agreement with those obtained using the CRYSTAL
program, where Ewald summation is used to perform the Coulomb 
lattice sums. 
In future, we intend to make our treatment of the Coulomb series more
efficient by incorporating real-space multipole expansion techniques.

The discrepancy between our Hartree-Fock results for all-{\em transoid}
polyethylene and the experimental ones was found to be most significant for 
the band structure. For example, the Hartree-Fock value of the direct band
gap is wrong by a factor of two as compared to the experimental one.
 These differences point to
the importance of electron-correlation effects. 
In a future
publication,  we will include these within
a local-correlation approach to study their influence on the  
quasi-particle properties of polyethylene and other insulating
systems\cite{graef1,graef2}.

\begin{table}  
 \protect\caption{ 
A summary of our HF results on all-{\em transoid} polyethylene
 performed with the 6-31G** 
basis set and its comparison with
the corresponding calculations performed by us with the CRYSTAL program
and the HF results of other authors. The Wannier-function-based 
calculations were performed with
up to fifth-nearest neighboring cells included in the orthogonality
region ($N=5$) and eighty five neighbors included in the Coulomb series
($M=85$). 
 Experimental values are also listed for
comparison. The lengths are expressed in the
units of $\AA$, the bond angles are in degrees, the total energy per C$_2$H$_4$
unit ($E_{\mbox{total}}$) is in Hartrees while the cohesive energy per CH$_2$ 
unit ($E_{\mbox{coh}}$) is in eV.}
 \protect\begin{center}  
  \begin{tabular}{ccccccc} \hline
         & $R_{CC}$ & $R_{CH}$ & $\alpha$ & $\beta$ & $E_{\mbox{total}}$ & 
         $E_{\mbox{coh}}$ \\
This work$^{a}$   &1.532 & 1.092 & 113.0  & 106.6     &-78.0728 & 9.85 \\
CRYSTAL           &1.533 & 1.092 & 113.2  & 106.5     &-78.0723 & 9.85 \\
Karpfen$^{b}$     &1.562 & 1.102 & 112.2  & 107.4     & ---     & ---  \\
Karpfen$^{c}$     &1.547 & 1.089 & 112.6  & 107.0     & ---     & ---  \\
Teramae et al.$^d$      
                  &1.565 & 1.102 & 115.3  & 104.7     & ---     & ---  \\
Exp.$^e$          &1.53  & 1.069 & 112.0  & 107.0     & ---     & ---  \\ 
Exp.$^f$          &1.578 & 1.06  & 107.7  & 109.0     & ---     & ---  \\ 
\hline
   \end{tabular}                      
   \end{center}  
$^a$ Performed with the lobe representation of the 6-31G** basis set. \\
$^b$ Performed with an extended basis set without polarization functions. 
Ref.~\protect\cite{pe-karp} \\
$^c$ Performed with the STO-3G basis set. Ref.~\protect\cite{pe-karp} \\
$^d$ Performed with the STO-3G basis set. Ref.~\protect\cite{pe-teramae} \\
$^e$ X-ray scattering results at 4 K. Refs. ~\protect\cite{pe-exp1,pe-exp2,pe-exp3} \\ 
$^f$ Neutron scattering results at 4 K. Ref. ~\protect\cite{pe-exp4} \\
  \label{tab-pef}    
\end{table}  
\end{document}